\documentclass[12pt,draftcls,lettersize,journal,onecolumn]{IEEEtran}
\usepackage{amsmath,amsfonts}
\usepackage{algorithm,algcompatible}
\usepackage{array}
\usepackage[caption=false,labelfont=sf,textfont=sf]{subfig}
\usepackage{textcomp}
\usepackage{stfloats}
\usepackage{url}
\usepackage{verbatim}
\usepackage{graphicx}
\usepackage{cite}
\usepackage{xcolor}
\graphicspath{{./figure/}}
\algnewcommand\INPUT{\item[\textbf{Input:}]}%
\algnewcommand\OUTPUT{\item[\textbf{Output:}]}%

\begin{document}

\title{Massive Data Generation for Deep Learning-aided Wireless Systems Using Meta Learning and Generative Adversarial Network}

\author{Jinhong~Kim,~Yongjun~Ahn,~and~Byonghyo~Shim
        % <-this % stops a space
\thanks{J. Kim, Y. Ahn, W. Kim, and B. Shim are with the Department of Electrical and Computer Engineering, Seoul National University, Seoul 08826, Korea (e-mail: {jhkim, yjahn, wjkim, bshim}@islab.snu.ac.kr).}
\thanks{This work was supported by Samsung Research Funding \& Incubation Center (SRFC-IT1901-17).}

\IEEEpubid{0000--0000/00\$00.00~\copyright~2021 IEEE}
}
\maketitle

\begin{abstract}
As an entirely-new paradigm to design the communication systems, deep learning (DL), an approach that the machine learns the desired wireless function, has received much attention recently.
In order to fully realize the benefit of DL-aided wireless system, we need to collect a large number of training samples.
Unfortunately, collecting massive samples in the real environments is very challenging since it requires significant signal transmission overhead.
In this paper, we propose a new type of data acquisition framework for DL-aided wireless systems.
In our work, generative adversarial network (GAN) is used to generate samples approximating the real samples.
To reduce the amount of training samples required for the wireless data generation, we train GAN with the help of the meta learning.
From numerical experiments, we show that the DL model trained by the GAN generated samples performs close to that trained by the real samples.
\end{abstract}


\section{Introduction}
\IEEEPARstart{I}{n} recent years, we have witnessed the emergence of artificial intelligence (AI)-based services such as driverless cars, smart factories, remote surgery, and drone-based delivery~\cite{zhang20196g, viswanathan2020communications}.
Communication mechanisms associated with these emerging applications and services are way different from traditional wireless systems in terms of latency, energy efficiency, reliability, and connection density.
As the wireless systems are becoming more complicated, it is very difficult to come up with a simple yet tractable mathematical model and algorithm.
As an entirely-new paradigm to handle future wireless systems, deep learning (DL), an approach that the machine learns the desired function without human intervention, has received much attention recently~\cite{liu2019deep, kim2020deep, ahn2021active, mei2021intelligent, jhkim2022parametric}.

Since the DL-based systems are data-driven in nature, to fully enjoy the benefit of DL-aided wireless system, sufficient training dataset is indispensable.
Unfortunately, collecting a large number of training samples in real-world wireless system is very difficult since it requires significant transmission overhead in terms of time, bandwidth, and power consumption.
For example, when one tries to collect one million received samples in 5G NR systems, it will take more than 15 minutes ($10^{6}$ symbols $\times$ 0.1 frame/symbol $\times$ 10 ms/frame).
To deal with the problem, one can use samples obtained by the mathematical channel model (e.g., extended pedestrian A (EPA) channel or extended vehicular A (EVA) channel model)~\cite{3GPP}.
Since the synthetic data can be generated using a simple computer programming, time and effort to collect huge training dataset can be greatly saved.
However, as the wireless channels are non-static in most cases and wireless environments are changing fast, a model mismatch caused by the variation of fading/noise/interference distribution and input statistics is unavoidable.
In such case, DL-based algorithm trained with a synthetic dataset would leave a considerable performance gap from the system using real data, resulting in a degradation of bit error rate (BER) and block error rate (BLER) performance.

An aim of this paper is to propose a new type of data acquisition framework for DL-aided wireless systems.
The key idea of the proposed strategy, dubbed as deep wireless data collection (D-WiDaC), is to acquire a massive number of real-like wireless samples using a \textit{generative adversarial network} (GAN).
In short, GAN is a DL model that generates samples approximating the input dataset~\cite{goodfellow2014generative}.
When GAN is trained properly, generated samples will be similar to the real samples, meaning that there is no fundamental difference between the GAN output and real samples.
Since the GAN training still requires a large amount of training samples, we exploit a meta learning, special training technique to quickly learn a task using a small number of samples~\cite{finn2017model}.
Since GAN pre-trained by the meta learning can exploit the common features in various wireless environments, it requires far smaller number of samples than that required by the vanilla (original) GAN.

From the simulations on the DL-based channel estimation in mmWave systems, we verify that the performance gap between the DL model trained by real channel dataset and that trained by D-WiDaC dataset is negligible.
Also, we demonstrate that the DL model trained by the D-WiDaC samples achieves more than 3 dB gain over that trained by the GAN-generated samples in terms of mean squared error (MSE).

The main contributions of this paper are as follows:
\begin{itemize}
    \item We design the CGAN-based DL architecture from which one can generate multiple wireless datasets associated with distinct characteristics (e.g., geometric condition, data traffic).
    In doing so, we greatly save the time and effort to obtain the real-like samples for various wireless systems.
    \item We design the two-stage CGAN training strategy consisting of the meta learning and fine-tuning.
    In the meta learning phase, common features in multiple wireless datasets are learned, and thus the role of the fine-tuning stage is to learn a few distinct features in the target environment.
    Since the fine-tuning requires a small number of samples, we can greatly reduce the CGAN training overhead.
    \item We numerically show the validity of the proposed scheme in various communication scenarios.
    In each scenario, we confirm that D-WiDaC saves more than 90\% of the samples to train the DL-based channel estimator.
\end{itemize}

The rest of this paper is organized as follows.
In Section II, we discuss GAN and the proposed D-WiDaC technique.
In Section III, we present the detailed methodologies to implement D-WiDaC to wireless applications.
In Section IV, we present the numerical results of the proposed method and conclude the paper in Section V.

\section{D-WiDaC for Wireless Data Collection}
\begin{figure}
    \begin{center}
        \includegraphics[width=\linewidth]{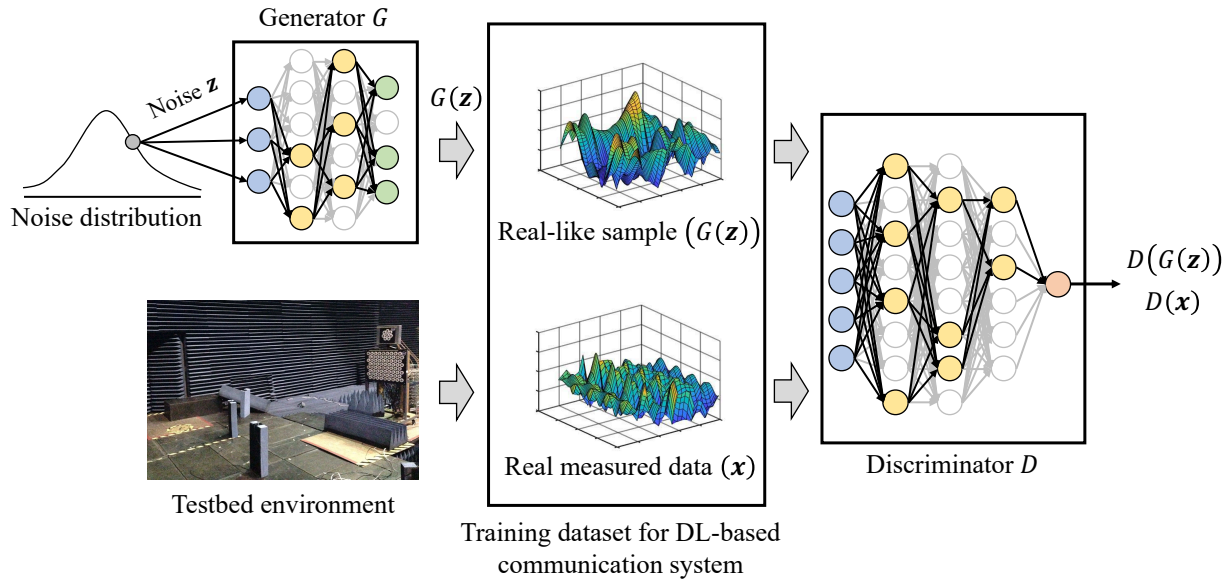}
    \end{center}
    \caption{Illustration of GAN-based data synthesis.}
    \label{fig:gan_data}
\end{figure}
In this section, we present the proposed D-WiDaC technique.
We first discuss the basics of GAN and then explain the D-WiDaC architecture and the meta learning-based training strategy.

\subsection{Basics of Generative Adversarial Network}

The main ingredients of GAN are a pair of DNNs called \textit{generator} $G$ and the \textit{discriminator} $D$.
The generator $G$ tries to produce the real-like data samples and the discriminator $D$ tries to distinguish real (authentic) and fake data samples.
To be specific, $G$ is trained to generate real-like data $G(\mathbf{z})$ from the random noise vector $\mathbf{z}$ and $D$ is trained to distinguish whether the generator output $G(\mathbf{z})$ is real or fake (see Fig.~\ref{fig:gan_data}).
In order to accomplish the mission, a min-max loss function, expressed as the cross-entropy\footnote{The cross-entropy between $x$ and $\hat{x}$ is defined as $H(x,\hat{x}) = - x \log(\hat{x}) - (1-x) \log(1- \hat{x})$.} between the distribution of generator output $G(\mathbf{z})$ and that of the real data $\mathbf{x}$, is used~\cite{goodfellow2014generative}:
\begin{align}
     \min_{G} \max_{D} \mathbb{E}_{\mathbf{x}}[\log(D(\mathbf{x}))]+ \mathbb{E}_{\mathbf{z}}[\log(1-D(G(\mathbf{z})))],
\end{align}
where $D(\mathbf{x})$ is the discriminator output which corresponds to the probability of $\mathbf{x}$ being real (non-fake).
In the training process, parameters of $G$ are updated while those of $D$ are fixed and vice versa.
When the training is finished properly, the generator output $G(\mathbf{z})$ is fairly reliable so that the discriminator cannot judge whether the generator output is real or fake (i.e., $D(G(\mathbf{z})) \approx 0.5$).
This means that we can safely use the generator output for the training purpose.

\subsection{D-WiDaC Architecture}
The key idea of D-WiDaC is to collect real-like wireless samples using GAN.
When collecting samples, we need to make sure that GAN generates wireless samples of interest since otherwise GAN might generate samples irrelevant to the desired wireless environment.
To do so, we use a special type of GAN, called conditional GAN (CGAN).
The distinct feature of CGAN over the vanilla GAN is to use an additional input on top of the random noise, called condition $\mathbf{c}$.
In essence, the condition $\mathbf{c}$ is an indicator (e.g., one-hot vector $[0 \ 1 \ 0 \ \cdots \ 0]$ or a scalar value) that points out the type of samples we want to generate.
In the proposed D-WiDaC, we design the condition such that it properly designates the target wireless environment.
For example, if we want to collect samples for the second channel among 5 distinct channels, we set $\mathbf{c} = \left[ 0 \ 1 \ 0 \ 0 \ 0\right]$.

\begin{figure}
	\begin{center}
		\includegraphics[width=0.8\linewidth]{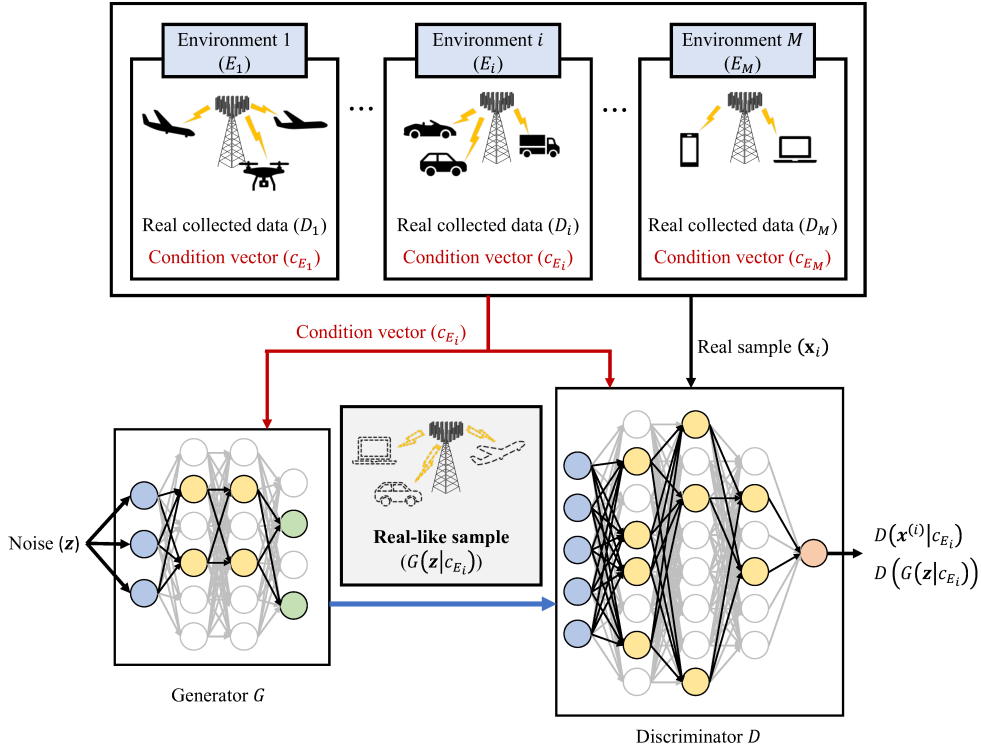}
	\end{center}
    \caption{D-WiDaC architecture.}
	\label{fig:D_WiDaC}
\end{figure}

To be specific, let $\mathbf{x}^{(i)}$ and $D_i = [\mathbf{x}^{(i,1)},\cdots,\mathbf{x}^{(i,N)}]$ ($i=1,\cdots,M$) be a real sample and the set of real samples of $i$-th dataset, respectively.
Also, let $\mathcal{L}_{D_i}$ and $\mathbf{c}_{i}$ be the loss function of CGAN and the condition corresponding to $D_{i}$, respectively.
Then, the loss function $\mathcal{L}_{D_i}$ is expressed as~\cite{mirza2014conditional}
\begin{align}
\mathcal{L}_{D_i} = \min_{G} \max_{D} \ &\mathbb{E}_{\mathbf{x}^{(i)}}[\log(D(\mathbf{x}^{(i)}|\mathbf{c}_{i}))]\\&+\mathbb{E}_{\mathbf{z}}[\log(1-D(G(\mathbf{z}|\mathbf{c}_{i})))].
\label{eq:maml_gan}
\end{align}
When CGAN is trained properly, it generates samples close to the target wireless environment (see Fig.~\ref{fig:D_WiDaC}).

\subsection{D-WiDaC Training}
As mentioned, the main goal of D-WiDaC is to generate massive real-like wireless samples with a small number of real samples.
In reality, however, CGAN still requires considerable training samples and hence the practical benefit of the proposed technique might be washed away.
To overcome the shortcoming, we exploit the \textit{meta learning}, a technique to train a model on a variety of tasks such that it can solve new task using only a small number of training samples~\cite{finn2017model}.
In short, meta learning is a special training technique to obtain the initialization parameters of DNN using which one can easily and quickly learn the desired function with a few training samples.

\begin{algorithm}[t!]
    \caption{Training Process of D-WiDaC}
  \begin{algorithmic}[1]
    \INPUT Wireless data $\{D_i\}^{M+1}_{i=1}$, condition $\{\mathbf{c}_i\}^{M+1}_{i=1}$, learning rates $\alpha$, $\beta$, and $\gamma$
    \STATE randomly initialize GAN parameters $\theta$
    \WHILE{meta learning}
        \FOR{$i \leftarrow 1$ to $M$}
        \STATE Sample batch data $\mathbf{x}^{(i)}$ from $D_i$
        \STATE Evaluate $\nabla_\theta \mathcal{L}_{D_i}$ using $\mathbf{x}^{(i)}$, $\mathbf{c}_i$, and CGAN loss $\mathcal{L}_{D_i}$ in Equation~(\ref{eq:maml_gan})
        \STATE Compute adapted parameters with gradient descent: $\psi_{D_i} = \theta - \alpha \nabla_\theta \mathcal{L}_{D_i}(\theta)$
        \STATE Sample batch data for meta-update $\mathbf{x'^{(i)}}$ from $D_i$
        \ENDFOR
    \STATE Update $\theta = \theta - \beta \nabla_\theta \sum_{i=1}^{M} \mathcal{L}_{D_i}(\psi_{D_i})$ using $\mathbf{x'^{(i)}}$, $\mathbf{c}_i$, and CGAN loss $\mathcal{L}_{D_i}$ in Equation~(\ref{eq:maml_gan})
    \ENDWHILE
    
    \WHILE{parameter update}
    \STATE Sample batch data $\mathbf{x}^{(M+1)}$ from $D_{M+1}$
        \STATE Evaluate $\nabla_\theta \mathcal{L}_{D_{M+1}}$ using $\mathbf{x}^{(M+1)}$, $\mathbf{c}_{M+1}$, and CGAN loss $\mathcal{L}_{D_{M+1}}$ in Equation~(\ref{eq:maml_gan})    
        \STATE Compute adapted parameters with gradient descent: $\theta = \theta - \gamma \nabla_\theta \mathcal{L}_{D_{M+1}}(\theta)$
    \ENDWHILE

  \end{algorithmic}
  \label{algorithm}
\end{algorithm}

Overall procedure of D-WiDaC training is as follows.
First, we perform the meta learning to obtain the initialization parameters.
We then update the network parameters to perform the fine-tuning of DNN such that the trained DNN generates samples for the desired wireless environments.
In the meta learning phase, we extract the common features of multiple wireless datasets, say $M$ datasets $\left\{ D_{1}, \cdots, D_{M}\right\}$, and then use them to obtain the network initialization parameters $\theta$:
\begin{align}
    \psi_{D_i, t} &= \theta_{t-1} - \alpha \nabla_\theta \mathcal{L}_{D_i}(\theta_{t-1}), \label{eq:individual_loss}\\
    \theta_{t} &= \theta_{t-1} - \beta \nabla_\theta \sum_{i=1}^{M} \mathcal{L}_{D_i}(\psi_{D_i,t}), \label{eq:meta_loss}
\end{align}
where $\theta_t$ and $\theta_{t-1}$ are the parameters updated by using $M$ datasets in $t$-th step and $(t-1)$-th step, respectively.
Also, $\psi_{D_i, t}$ is the parameter associated with dataset $D_i$ in $t$-th step, $\mathcal{L}_{D_i}$ is the loss function of CGAN for $i$-th dataset $D_i$, and $\alpha$ and $\beta$ are the step sizes for the parameter update (see Algorithm~\ref{algorithm}).
In each iteration, we temporarily update the CGAN parameters for each dataset to obtain $\{ \psi_{D_{i},t}\}_{i=1}^{M}$ (see~\eqref{eq:individual_loss}).
We then update the CGAN parameters $\theta$ in the direction to minimize the sum of losses for $\{ \psi_{D_{i},t}\}_{i=1}^{M}$ (see~\eqref{eq:meta_loss}).
In doing so, the CGAN parameters $\theta$ learn the common features in the $M$ datasets $\{D_{i}\}_{i=1}^{M}$.
Then, in the fine-tuning phase, we use $\theta$ as an initialization point of D-WiDaC.
Since all we need in the fine-tuning is to learn the distinct features (of $D_{M+1}$) unextracted from the meta learning, we can greatly reduce the overhead to collect $D_{M+1}$ samples.

To show the effectiveness of the meta learning in the proposed D-WiDaC, we briefly explain the following analytic argument.
In~\cite{fallah2021generalization}, it has been shown that the gap between the losses for datasets $\{D_{i}\}_{i=1}^M$ and a new dataset $D_{M+1}$ is bounded after the meta learning.
To be specific, the gap between the losses $\mathcal{L}_{D_{M+1}}(\theta^{\ast})$ and $\frac{1}{M}\sum^{M}_{i=1}\mathcal{L}_{D_{i}}(\theta^{\ast})$ is smaller than the \textit{total variation distance} $f(D_{M+1}, \{ D_i \}_{i=1}^M)$ which measures the difference between two distributions $P(D_{M+1})$ and $P(\{ D_i \}_{i=1}^M)$:
\begin{align}
   \vert \mathcal{L}_{D_{M+1}}(\theta^{\ast})-\frac{1}{M}\sum^{M}_{i=1}\mathcal{L}_{D_{i}}(\theta^{\ast}) \vert \leq f(D_{M+1}, \{ D_i \}_{i=1}^M),
   \label{eq:analysis}
\end{align}
where $\theta^{\ast}$ is the optimal parameters minimizing $\frac{1}{M}\sum^{M}_{i=1}\mathcal{L}_{D_{i}}(\theta^{\ast})$.

In many communication scenarios, distributions of the wireless datasets are more or less similar since the key characteristics of the wireless channels remain unchanged except for a few distinct ones.
In our context, this directly implies that the parameters $\theta^{\ast}$ obtained via meta learning would be very close to the optimal CGAN parameter for $D_{M+1}$.
Thus, by using the meta-trained CGAN parameters $\theta$ as an initialization parameters, we can accelerate CGAN training in the newly observed wireless environment.

\subsection{D-WiDaC Implementation Example}
In this subsection, we explain the wireless channel sample generation using the proposed D-WiDaC technique.

As an example, we consider a narrowband geometric channel model with MISO system where the numbers of transmit antennas and receive antenna are $N_t$ and 1, respectively.
In this setup, the propagation channel model $\mathbf{h}\in \mathbb{C}^{N_{t} \times 1}$ between the transmitter and receiver can be expressed as\footnote{We simply let the antenna array of the transmitter be the uniform linear array (ULA).}
\begin{align}
    \label{eq:channel_model} \mathbf{h} &= \sqrt{\frac{N_{t}}{L}} \sum_{l=1}^{L} \rho_{l} \mathbf{a}(\theta_{l}), 
\end{align}
\begin{align}
    \rho_{l} &\sim \mathcal{CN}(0,C) = \mathcal{CN}(0,\frac{P_{0}}{f^{2} R^{2}}), \label{eq:7}\\
    \mathbf{a}(\theta_{l}) &= \frac{1}{N_{t}} \left[ 1, e^{j\theta_{l}},\cdots,e^{j (N_{t}-1)\theta_{l}} \right]^{T}, \label{eq:8}
\end{align}
where $\rho_{l}$, $C$, $P_{0}$, $f$, $R$, $\theta_{l}$, $\mathbf{a}$, and $L$ are the complex gain, distance-dependent path loss, power gain, center frequency, distance, azimuth angle of departure (AoD), transmit array response vector associated with the $l$-th propagation path, and the number of paths, respectively.

When we try to generate the channel samples without the channel information described in~\eqref{eq:channel_model}-\eqref{eq:8}, we apply D-WiDaC as follows.
Let $D_1$, $D_2$, $D_3$, and $D_4$ be the channel dataset at the central frequency $f=0.9,\,3.5,\,28$, and $60\,\text{GHz}$.
As an input of the generator $G$, we use the concatenation of the random noise vector $\mathbf{z}$ and condition $\mathbf{c}$.
We assume that sufficient number of real samples for $f=0.9,\,3.5,$ and $60\,\text{GHz}$ channels are available but not for $f=28\,\text{GHz}$ channel.

In the meta learning phase, we use $\left\{\mathbf{c}_{1},D_{1}\right\}$, $\left\{\mathbf{c}_{2},D_{2}\right\}$, and $\left\{\mathbf{c}_{4},D_{4}\right\}$ to extract the common features such as the number of dominant paths, AoD distribution, and transmit array response.
When the meta learning is finished, we perform the fine-tuning of D-WiDaC parameters using $\left\{\mathbf{c}_{3}, D_{3}\right\}$ to generate the channels corresponding to $f=28\,\text{GHz}$.
Since the mission of D-WiDaC in this parameter update phase is to learn the unique features of $28\,\text{GHz}$ channel, we can train the network with small number of training samples.

\section{Simulation Results}

\begin{figure*}
     \centering
     \subfloat[]{
        \includegraphics[width=0.33\linewidth]{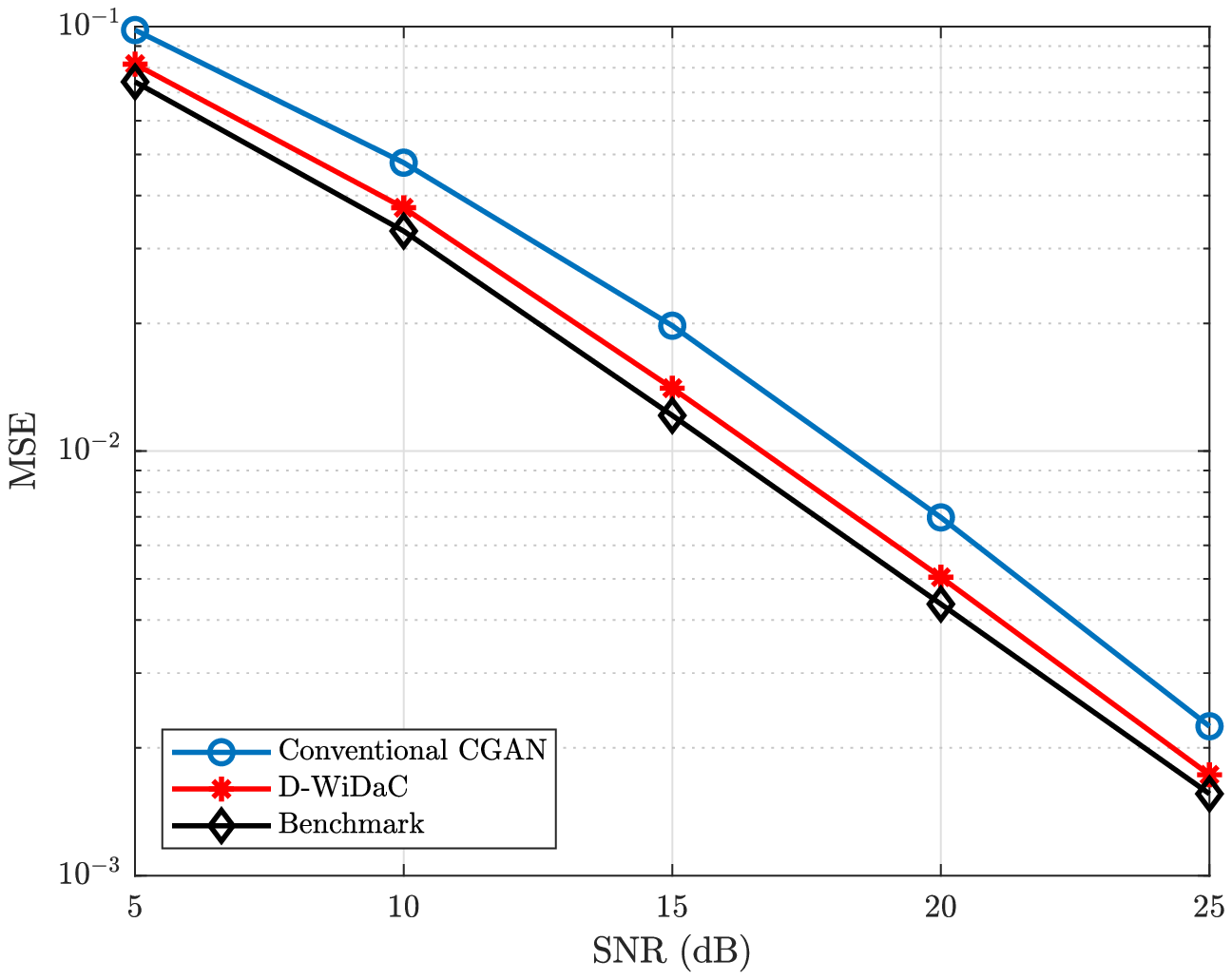}
     }
     \subfloat[]{
        \includegraphics[width=0.33\linewidth]{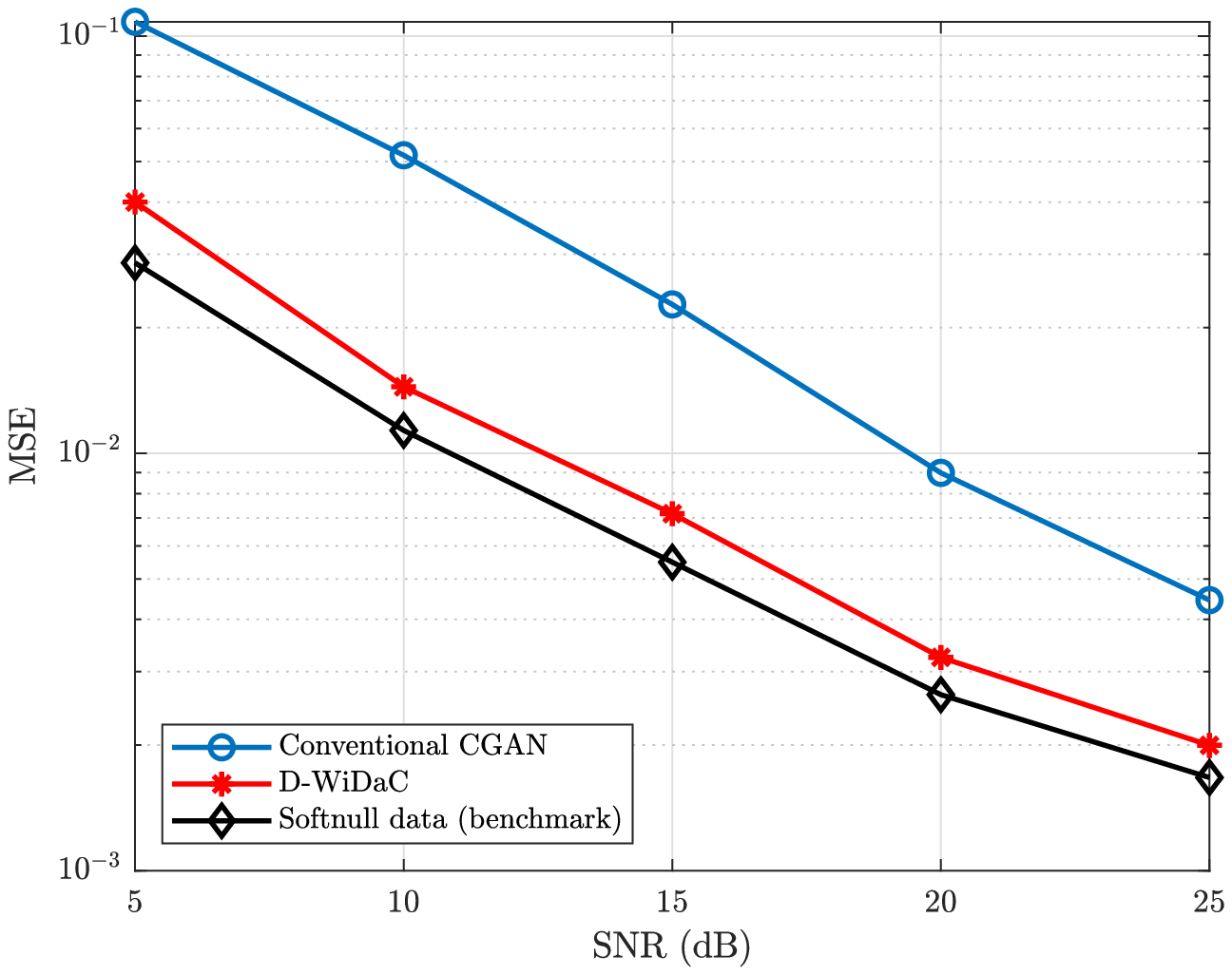}
     }
     \subfloat[]{
        \includegraphics[width=0.33\linewidth]{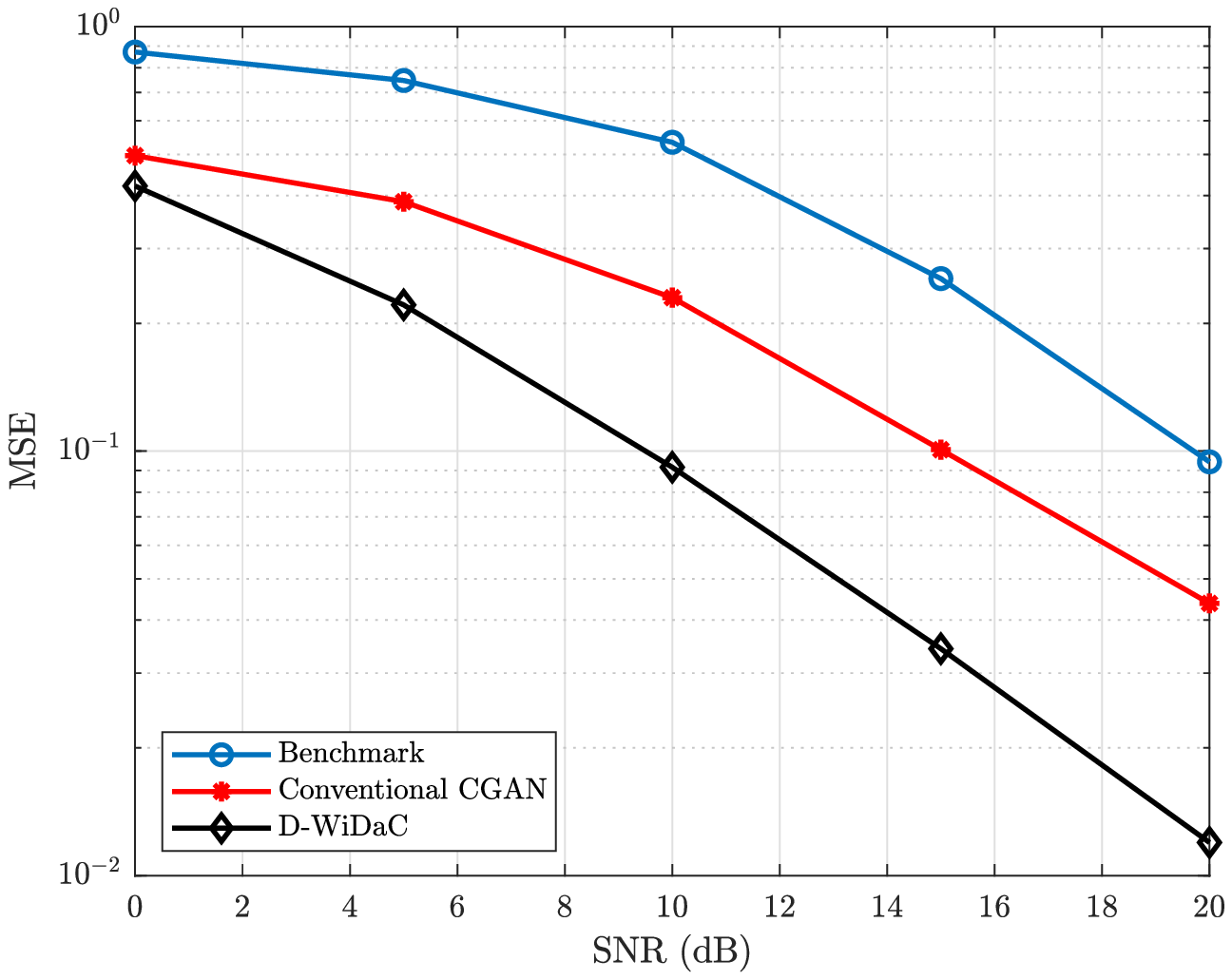}
     }
    \caption{MSE performance of the DL-based channel estimator using three distinct datasets: genie dataset, generated dataset from conventional CGAN and D-WiDaC. (a) Model-based channel samples. We use 10,000 samples of 39 GHz channel for testing. (b) Real measured (softnull) dataset. We use 2,500 samples of 5 ft channel for testing. (c) IRS-aided system~\cite{huang2019reconfigurable} channel samples. We use 10,000 samples for 4 different BS to IRS link distances; 5, 10, 15, and 20$\,$m.}
    \label{fig:MSE_comparison}
\end{figure*}

In order to observe the validity of the proposed data acquisition strategy, we evaluate the MSE performance of the DL-based channel estimator\footnote{As a DL-based channel estimator, we use fully-connected network consisting of 5 hidden layers, each of which has 256 hidden units. Also, in the training process, we simply use the channel MSE as a loss function.} trained by the samples generated by D-WiDaC.
Specifically, to investigate the efficacy of D-WiDaC, we use two different types of benchmark datasets: model-based channel samples and real measured channel samples.
As a model-based channel dataset, we exploit the samples generated from (\ref{eq:channel_model}).
As a measured channel dataset, we employ the softnull dataset obtained by massive MIMO systems at indoor environments~\cite{SoftNull}.

In Fig.~\ref{fig:MSE_comparison} (a), we investigate the MSE performance of DL-based channel estimator trained by three different training datasets: 1) dataset obtained from (\ref{eq:channel_model}) (we call it genie dataset), 2) dataset generated from conventional CGAN (without meta learning), and 3) dataset generated from D-WiDaC.
For the meta learning of D-WiDaC, we exploit 80,000 samples corresponding to 28, 37, 41, and 60 GHz channel ($M=4$).
Also, we use 800 samples of 39 GHz channel for the training of conventional CGAN and the D-WiDaC fine-tuning.
For the training of the DL-based channel estimation at 39 GHz, we use 200,000 samples for all techniques under test. 

As shown in Fig.~\ref{fig:MSE_comparison} (a), the MSE performances of the DL models trained by the benchmark dataset and D-WiDaC dataset are more or less similar since D-WiDaC trained by using various channel dataset can well extract the common channel features.
In fact, D-WiDaC can significantly reduce the number of real samples for the DL model training (in our case, $\frac{200,000-800}{200,000} = 99.6\%$ reduction of 39 GHz channel samples) as long as multiple wireless datasets having common features are available.
Whereas, the DL model trained by the conventional CGAN-based samples performs poor (around 3 dB loss at MSE=$10^{-2}$) since the number of training samples is not sufficient enough to train the generator $G$ and discriminator $D$ of CGAN.

\begin{figure*}
     \centering
     \subfloat[]{
        \includegraphics[width=0.5\linewidth]{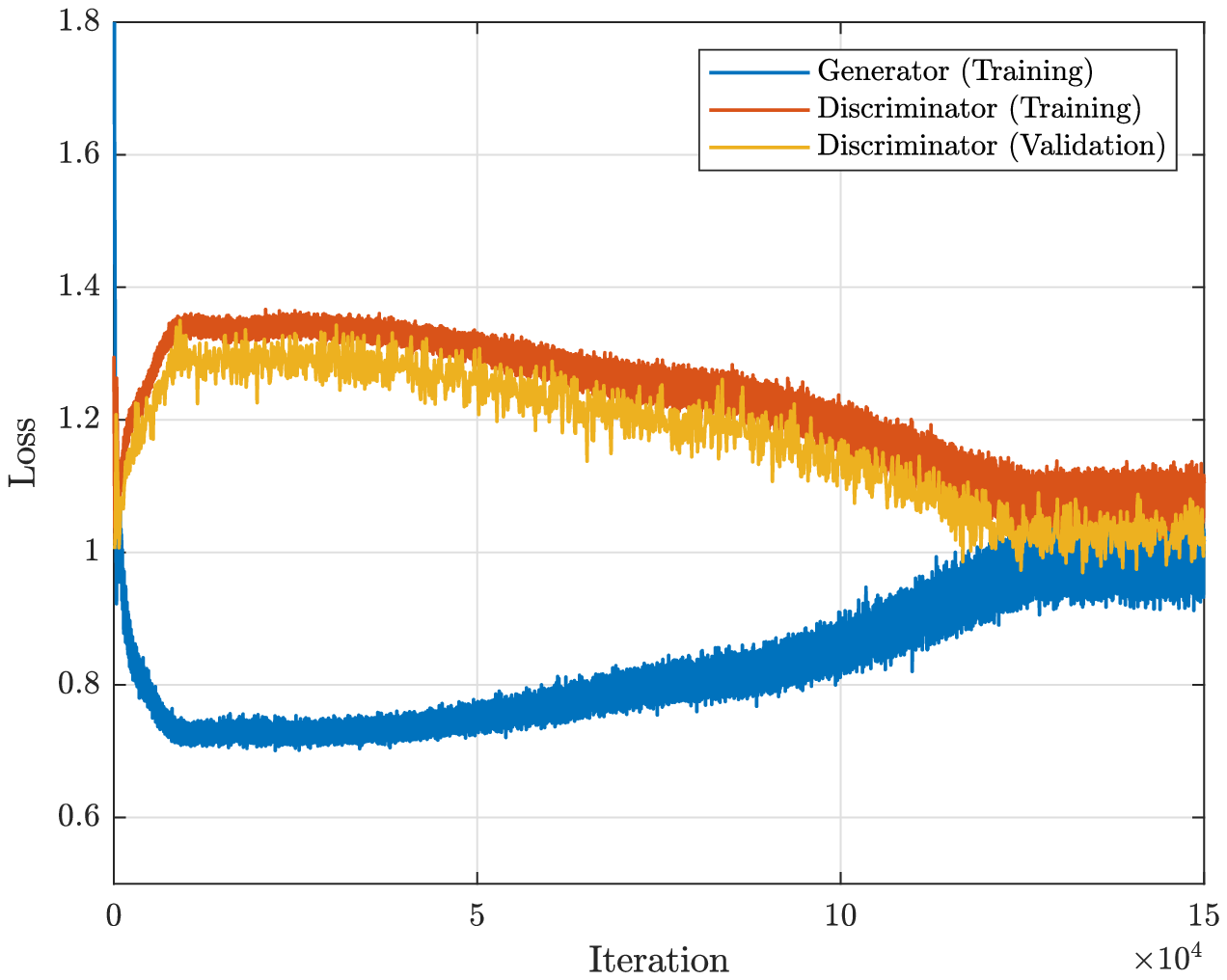}
     }
     \subfloat[]{
        \includegraphics[width=0.5\linewidth]{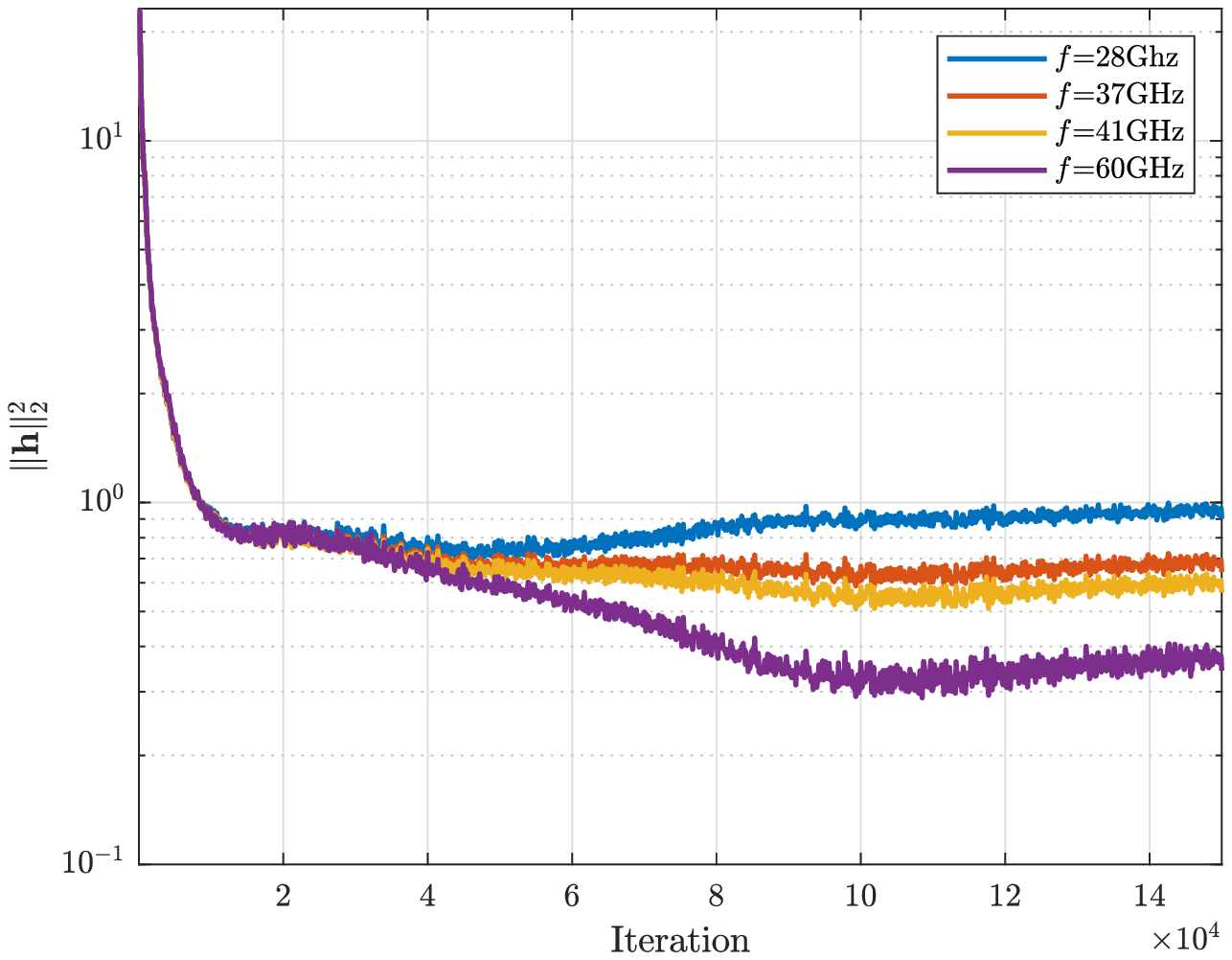}
     }
        \caption{CGAN model evaluations as a function of training iterations in the meta learning phase: (a) training and validation losses and (b) path gain of model-based channel samples for various center frequencies. We generate the channel data for $f=28,37,41$, and $60\,$GHz in every 100 iterations.}
        \label{fig:CGAN_training}
\end{figure*}

We next evaluate the performance of D-WiDaC for the real channel samples~\cite{SoftNull}.
In this test, we use the samples characterized by distinct propagation distances (3, 4, 5, 6, and 7 ft).
For the meta learning of D-WiDaC, we use 2,600 samples corresponding to 3, 4, 6, and 7 ft channel ($M=4$).
Also, we use 1,000 samples of 5 ft channel for the training of conventional CGAN and the D-WiDaC fine-tuning.
In the training of the DL-based channel estimator, we use 10,000 samples of 5 ft channel.

In Fig.~\ref{fig:MSE_comparison} (b), we test the MSE performances of the DL-based channel estimator trained by three different datasets: softnull dataset, dataset generated from the CGAN, and the proposed D-WiDaC.
We observe that the channel estimation performance of the D-WiDaC-based approach is slightly worse than that using the real samples (e.g., 1.7 dB loss at MSE=$10^{-2}$).
Whereas, the performance gap of the CGAN-based approach and the case using real samples is large (more than 6 dB at MSE=$10^{-2}$) since this approach does not have a mechanism to exploit the common features of diverse wireless environments.

To validate the effectiveness of D-WiDaC in the complex wireless systems, we plot the MSE performance of the DL-based intelligent reflecting surface (IRS) channel estimation (see Fig.~\ref{fig:MSE_comparison} (c)).
In Fig.~\ref{fig:MSE_comparison} (c), we observe that the performance of the DL-based channel estimator using D-WiDaC samples is close to that of the DL model trained by the benchmark dataset.
From this result, we see that the proposed D-WiDaC can also reduce the data collection overhead required for the RIS channel measurement campaign.

In Fig.~\ref{fig:CGAN_training} (a), we evaluate the training and validation errors of D-WiDaC.
To quantify the training error, we measure the generator and discriminator losses for the training dataset.
From the experiments, we observe that the training loss of D-WiDaC converges after 130,000 iterations.
During the training, we also measure the discriminator loss using the validation dataset.
We see that the validation loss converges without suffering from overfitting since the meta learning provides sufficient amount of multiple datasets.

To see if D-WiDaC properly generates the samples for the target environment, we measure the path gain of the model-based channel samples on various center frequencies including $f=28,37,41$, and $60\,$GHz (see Fig.~\ref{fig:CGAN_training} (b)).
We observe that the path gain of the D-WiDaC channels converges to that of the model-based channels for all center frequencies $f=28,37,41$, and $60\,$GHz as the training iteration increases.
For example, path gain of $f=$28$\,$GHz channel data is 0.98 after 100,000 iterations, which is almost the same as that of the benchmark, 1.04.

\begin{table}[]
\centering
\caption{Data generation overhead comparison with the conventional techniques. For the conventional data generation techniques, we set the total number of data samples is 200,000.}
\label{tab:tabel_1}
{\small
\begin{tabular}{|l|r|}
\hline
        & \multicolumn{1}{c|}{FLOPs}\\ \hline
D-WiDaC & 143,896     \\ \hline
SMOTE   & 7,288,540   \\ \hline
MSMOTE  & 161,035,040 \\ \hline
INOS    & 204,132,150 \\ \hline
\end{tabular}
}
\vspace{-1em}
\end{table}

In Table~\ref{tab:tabel_1}, we verify the data generation overhead of the D-WiDaC.
For comparison, we measure the number of floating point operations (flops) of D-WiDaC and conventional data generation techniques including 1) SMOTE, 2) MSMOTE, and 3) INOS.
Specifically, we measure the number of flops required to generate one channel sample\footnote{In our simulation, we use 3 FC layers in the generator layers, each of which uses 256 elements.
Also, we set the input and output vector dimensions as $\mathbf{z} \in \mathbb{R}^{8 \times 1}$, $c \in \mathbb{R}$, and $N_t = 8$.}.
As shown in Table~\ref{tab:tabel_1}, the number of flops of the proposed D-WiDaC is smaller than conventional data generation techniques.
Since the D-WiDaC only uses a few steps of simple multiplications and additions, the data generation overhead is far smaller than the conventional techniques requiring complicated sorting before the data generation.

In Table~\ref{tab:tabel_2}, we summarize the path gain of the generated samples to show the efficacy of the fine-tuning process.
To be specific, we measure the average path gains of the $f=39\,\text{GHz}$ channel data samples generated (with and without fine-tuning).
We observe that the path gain of samples from the fine-tuned model is closer to the path gain of samples obtained from the meta learning only.
This directly means that the fine-tuning process enhances the data generation performance with a few training samples.

\begin{table}[]
\centering
\caption{Average path gain of generated data samples in fine-tuned model and meta trained model}
\label{tab:tabel_2}
{\small
\begin{tabular}{|l|c|}
\hline
                     & Average path gain \\ \hline
Benchmark dataset    & 0.561              \\ \hline
Fine-tuned dataset   & 0.583              \\ \hline
Meta learned dataset & 0.644              \\ \hline
\end{tabular}
}
\vspace{-1em}
\end{table}

\section{Conclusion}
In this paper, we proposed a new type of wireless data acquisition framework for the DL-aided wireless systems.
The key idea behind the proposed D-WiDaC technique is to exploit CGAN and meta learning to reduce the training sample overhead.
We demonstrated from the numerical evaluations that the proposed scheme is effective in generating the realistic wireless data and reducing the number of real samples over the vanilla CGAN training.
There are many wireless datasets that slightly differ in some characteristics (e.g., center frequency, propagation distance, level of interference, and Doppler frequency).
If we properly design the condition $\mathbf{c}$ that can describe these various features and perform meta learning, the problem caused by the lack of samples will be greatly alleviated.
We expect that our meta learning-based approach will be more effective in the 6G era where the datasets generated from similar but distinct wireless environments will be sufficient.
For the test code of wireless examples discussed in this paper, check out http://islab.snu.ac.kr/publication.

\ifCLASSOPTIONcaptionsoff
  \newpage
\fi

\bibliographystyle{IEEEtran}
\bibliography{refs}

\begin{thebibliography}{10}
\providecommand{\url}[1]{#1}
\csname url@samestyle\endcsname
\providecommand{\newblock}{\relax}
\providecommand{\bibinfo}[2]{#2}
\providecommand{\BIBentrySTDinterwordspacing}{\spaceskip=0pt\relax}
\providecommand{\BIBentryALTinterwordstretchfactor}{4}
\providecommand{\BIBentryALTinterwordspacing}{\spaceskip=\fontdimen2\font plus
\BIBentryALTinterwordstretchfactor\fontdimen3\font minus
  \fontdimen4\font\relax}
\providecommand{\BIBforeignlanguage}[2]{{%
\expandafter\ifx\csname l@#1\endcsname\relax
\typeout{** WARNING: IEEEtran.bst: No hyphenation pattern has been}%
\typeout{** loaded for the language `#1'. Using the pattern for}%
\typeout{** the default language instead.}%
\else
\language=\csname l@#1\endcsname
\fi
#2}}
\providecommand{\BIBdecl}{\relax}
\BIBdecl

\bibitem{zhang20196g}
Z.~Zhang, Y.~Xiao, Z.~Ma, M.~Xiao, Z.~Ding, X.~Lei, G.~K. Karagiannidis, and
  P.~Fan, ``{6G} wireless networks: {V}ision, requirements, architecture, and
  key technologies,'' \emph{IEEE Veh. Technol. Mag.}, vol.~14, no.~3, pp.
  28--41, Sep. 2019.

\bibitem{viswanathan2020communications}
H.~Viswanathan and P.~E. Mogensen, ``Communications in the 6{G} era,''
  \emph{IEEE Access}, vol.~8, pp. 57\,063--57\,074, Mar. 2020.

\bibitem{liu2019deep}
Y.~Liu, H.~Yu, S.~Xie, and Y.~Zhang, ``Deep reinforcement learning for
  offloading and resource allocation in vehicle edge computing and networks,''
  \emph{IEEE Trans. Veh. Technol.}, vol.~68, no.~11, pp. 11\,158--11\,168, Aug.
  2019.

\bibitem{kim2020deep}
W.~Kim, Y.~Ahn, and B.~Shim, ``{Deep neural network-based active user detection
  for grant-free NOMA systems},'' \emph{IEEE Trans. Commun.}, vol.~68, no.~4,
  pp. 2143--2155, Jan. 2020.

\bibitem{ahn2021active}
Y.~Ahn, W.~Kim, and B.~Shim, ``Active {U}ser {D}etection and {C}hannel
  {E}stimation for {M}assive {M}achine-{T}ype {C}ommunication: {D}eep
  {L}earning {A}pproach,'' \emph{IEEE Internet Things J.}, early access, Dec.
  2021.

\bibitem{mei2021intelligent}
J.~Mei, X.~Wang, K.~Zheng, G.~Boudreau, A.~B. Sediq, and H.~Abou-zeid,
  ``Intelligent {R}adio {A}ccess {N}etwork {S}licing for {S}ervice
  {P}rovisioning in 6{G}: {A} {H}ierarchical {D}eep {R}einforcement {L}earning
  {A}pproach,'' \emph{IEEE Trans. Commun.}, Jun. 2021.

\bibitem{jhkim2022parametric}
J.~Kim, Y.~Ahn, S.~Kim, and B.~Shim, ``Parametric sparse channel estimation
  using long short-term memory for mmwave massive mimo systems,'' in
  \emph{Proc. IEEE Int. Conf. Commun. (ICC)}, 2022.

\bibitem{3GPP}
\text{3GPP}~\text{TS} \text{36.104}, ``Evolved {U}niversal {T}errestrial
  {R}adio {A}ccess ({E-UTRA}); {B}ase {S}tation ({BS}) {R}adio {T}ransmission
  and {R}eception,'' \emph{3GPP; {T}echnical {S}pecification {G}roup {R}adio
  {A}ccess {N}etwork}.

\bibitem{goodfellow2014generative}
I.~Goodfellow, J.~Pouget-Abadie, M.~Mirza, B.~Xu, D.~Warde-Farley, S.~Ozair,
  A.~Courville, and Y.~Bengio, ``Generative adversarial nets,'' \emph{Adv.
  Neural Inf. Proc. Syst.}, vol.~27, 2014.

\bibitem{finn2017model}
C.~Finn, P.~Abbeel, and S.~Levine, ``Model-agnostic meta-learning for fast
  adaptation of deep networks,'' in \emph{Proc. Int. Conf. Mach. Learn.
  (ICML)}.\hskip 1em plus 0.5em minus 0.4em\relax PMLR, Aug. 2017, pp.
  1126--1135.

\bibitem{mirza2014conditional}
M.~Mirza and S.~Osindero, ``Conditional generative adversarial nets,''
  \emph{arXiv preprint arXiv:1411.1784}, 2014.

\bibitem{fallah2021generalization}
A.~Fallah, A.~Mokhtari, and A.~Ozdaglar, ``Generalization of model-agnostic
  meta-learning algorithms: Recurring and unseen tasks,'' \emph{Adv. Neural
  Inf. Process. Syst.}, vol.~34, 2021.

\bibitem{huang2019reconfigurable}
C.~Huang, A.~Zappone, G.~C. Alexandropoulos, M.~Debbah, and C.~Yuen,
  ``Reconfigurable intelligent surfaces for energy efficiency in wireless
  communication,'' \emph{IEEE Trans. Wireless Commun.}, vol.~18, no.~8, pp.
  4157--4170, 2019.

\bibitem{SoftNull}
E.~Everett, C.~Shepard, L.~Zhong, and A.~Sabharwal, ``Softnull: Many-antenna
  full-duplex wireless via digital beamforming,'' \emph{IEEE Trans. Wireless
  Commun.}, vol.~15, no.~12, p. 8077 — 8092, Dec. 2016.

\end{thebibliography}

\end{document}